\documentclass[%
preprint,
 amsmath,amssymb,
 aps
]{revtex4}

\usepackage{graphicx}
\usepackage{dcolumn}
\usepackage{bm}

\begin{document}


\title{Characterization of spin-orbit fields in InGaAs quantum wells}


\author{T. Henn}
\affiliation{IBM Research - Z{\"u}rich, S{\"a}umerstrasse 4, 8803 R{\"u}schlikon, Switzerland}

\author{L. Czornomaz}
\affiliation{IBM Research - Z{\"u}rich, S{\"a}umerstrasse 4, 8803 R{\"u}schlikon, Switzerland}

\author{G. Salis}
\email[]{gsa@zurich.ibm.com}
\affiliation{IBM Research - Z{\"u}rich, S{\"a}umerstrasse 4, 8803 R{\"u}schlikon, Switzerland}

\date{\today}

\begin{abstract}
Coherent electron spin dynamics in 10-nm-wide InGaAs/InAlAs quantum wells is studied from 10\,K to room temperature using time-resolved Kerr rotation. The spin lifetime exceeds 1\,ns at 10\,K and decreases with temperature. By varying the spatial overlap between pump and probe pulses, a diffusive velocity is imprinted on the measured electron spins and a spin precession in the spin-orbit field is measured. A Rashba symmetry of the SOI is determined. By comparing the spatial precession frequency gradient with the spin decay rate, an upper limit for the Rashba coefficients $\alpha$ of 2$\times$10$^{-12}$\,eVm is estimated.
\end{abstract}

\maketitle
Spin-orbit interaction (SOI) couples the electron's spin to its motion and is of great significance for spin manipulation by electric fields and currents, as proposed and implemented in spintronics devices~\cite{Zutic2004,Sinova2015} and spin-based quantum bits~\cite{Nowack2007}. Much interest has recently been shown in using materials with large SOI to implement helical modes that are spin-polarized and counterpropagating, and that are expected to be protected against elastic back-scattering~\cite{Qi2011}. Such systems are also important for Majorana zero modes~\cite{Sau2010,Oreg2010} and generally require semiconductor materials with small band gaps, such as InAs or InSb. The Rashba-type SOI~\cite{Bychkov1984}, especially, increases dramatically when decreasing the band-gap of the semiconductor.

SOI in two-dimensional electron gases confined in semiconductor quantum-well (QW) structures can be measured using transport or optical techniques. Time-resolved optical pump-probe methods~\cite{Meier2007,Koralek2009,Kohda2015} have proven especially useful to directly access the SOI and have thereby provided rich information about the size and symmetry of SOI~\cite{Altmann2015, Altmann2016}. Up to now, such optical investigations were limited to the wavelength range below 1000\,nm, based on the convenience of available lasers and detectors. To investigate materials with larger SOI, an extension of such techniques to larger wavelengths is desired.

In Ref.~\onlinecite{Kohda2015}, an all-optical technique based on imprinting a diffusive spin motion was developed to measure SOI in GaAs quantum wells. Here, we adapt this technique to the 1500\,nm telecom wavelength region. Investigating the coherent electron spin dynamics in In$_{0.53}$Ga$_{0.47}$As QWs placed in the intrinsic region of a n-i-p structure, we find spin decay times of 1.2\,ns at 10\,K and 120\,ps at room temperature. SOI is found to be dominated by the Rashba type. By comparing the variation of spin precession frequency versus pump-probe overlap position with the spin decay rate, we estimate an upper limit for the Rashba SOI coefficient $\alpha$ of 2$\times$10$^{-12}$\,eVm. In terms of energy splitting, this corresponds to 400\,$\mu$eV for a typical Fermi wavenumber of 1$\times$10$^8$\,m$^{-1}$.

The sample studied contains three 10-nm-wide In$_{0.53}$Ga$_{0.47}$As QWs, each embedded in 20-nm-wide In$_{0.52}$Al$_{0.48}$As barriers. The QWs are placed in between a 200-nm-thick n-doped region on the surface side and an equally thick p-doped region on the substrate side. The n- and p-doped regions were each divided into 100\,nm of lower doping concentration (2$\times$10$^{17}$\,cm$^{-3}$) on the QW side, and higher doping concentration (1$\times$10$^{19}$\,cm$^{-3}$) on the outer side. The sample was grown lattice-matched by metal-organic chemical vapor deposition on an InP substrate. The opposite doping on the two sides generates an electric field across the three QWs of about 8\,V/$\mu$m which is expected to establish a large Rashba SOI for the QW electrons.

The electron spin dynamics was measured using time-resolved Kerr rotation (TRKR)~\cite{Lau2006}. Pulses of tunable wavelength $\lambda$ were generated by an optical parametric oscillator (APE Berlin) pumped by a mode-locked Ti:sapphire laser at a repetition rate of 80\,MHz. After passing through a single-mode fiber, the pulse length was on the order of a few picoseconds. Pump pulses were circularly polarized and focused onto the sample surface (sigma-width of spot size 10 - 12\,$\mu$m, time-averaged power 2\,mW). Linearly polarized and time-delayed probe pulses (power 200\,$\mu$W, time delay $t$) were focused to a similar size. By choosing $\lambda$ close to the absorption edge of the QW, the circularly polarized pump pulses excite spin-polarized electrons into the conduction band of the QW layer, while a maximum Kerr rotation of the linearly polarized probe pulses is achieved. The Kerr rotation is measured using a balanced InGaAs photodiode bridge and is proportional to  the transient out-of-plane component of the electron spin polarization at time $t$. Using a motorized mirror, the relative spatial displacement between pump and probe spots was controlled along both dimensions of the sample surface. The sample was placed in a cryostat at temperatures between 10 and 296\,K. A magnetic field $B$ of up to 1\,T was applied in the plane of the sample.

\begin{figure}[t]
 \advance\leftskip0cm
\includegraphics[width=0.48\textwidth]{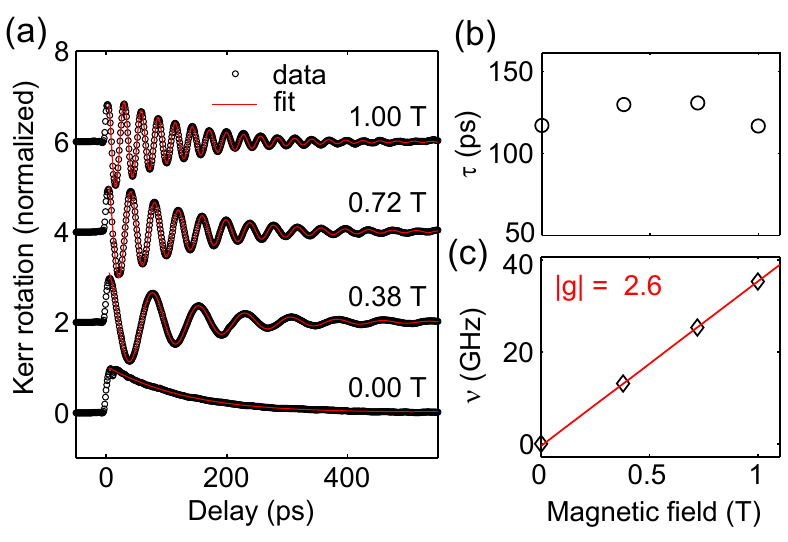}
\caption{\label{Fig1} (Color online). (a) Time-resolved Kerr rotation measurements (symbols) taken at room temperature track the dynamics of the optically excited electron spins at four different magnetic fields $B$. Red lines are fits to an exponentially decaying cosine oscillation. (b) The fitted decay times $\tau$ do not depend significantly on the applied magnetic field. (c) The fitted precession frequency $\nu$ increases linearly with $B$, with the slope indicating an electron g-factor of 2.6.}
\end{figure}

Data of TRKR at room temperature is shown in Fig.~\ref{Fig1}(a). At $B=1$\,T, the optically excited spin polarization precesses with a frequency $\nu$ close to 40\,GHz and decays within a time $\tau$ of a little more than 100\,ps. At $B=0$\,T, no spin precession is observed and the signal decays within a similar time constant. The data is fit (red curves) by an exponentially decaying cosine function, $a \exp(-t/\tau)\cos(2\pi\nu t)$, to obtain the decay time $\tau$, the precession frequency $\nu$, and the signal amplitude $a$. Time-resolved differential reflectivity measurements (not shown) indicate a charge recombination time much longer than 1\,ns in the whole temperature range, such that $\tau$ represents the lifetime of the electron spin. For magnetic fields $B$ between 0 and 1\,T, we observe a $\tau$ of around 120\,ps, approximately independent of $B$ [Fig.~\ref{Fig1}(b)]. This indicates that $\tau$ is not limited by inhomogeneities of spins probed within the extension of the laser spots like, e.g. a distribution of $g$ factors. Also, the high quality of the single-frequency fit indicates that the $g$-factors of the electrons measured in the three QWs are not significantly different. As expected, $\nu$ increases linearly with $B$ [Fig.~\ref{Fig1}(c)]. From the slope (red curve) we determine an electron g-factor of -2.6. The negative sign is inferred from the theoretical model in which band-gap dependent corrections of the g-factor are subtracted from a positive value of 2~\cite{Roth1959}.

\begin{figure}[t]
\includegraphics[width=0.48\textwidth]{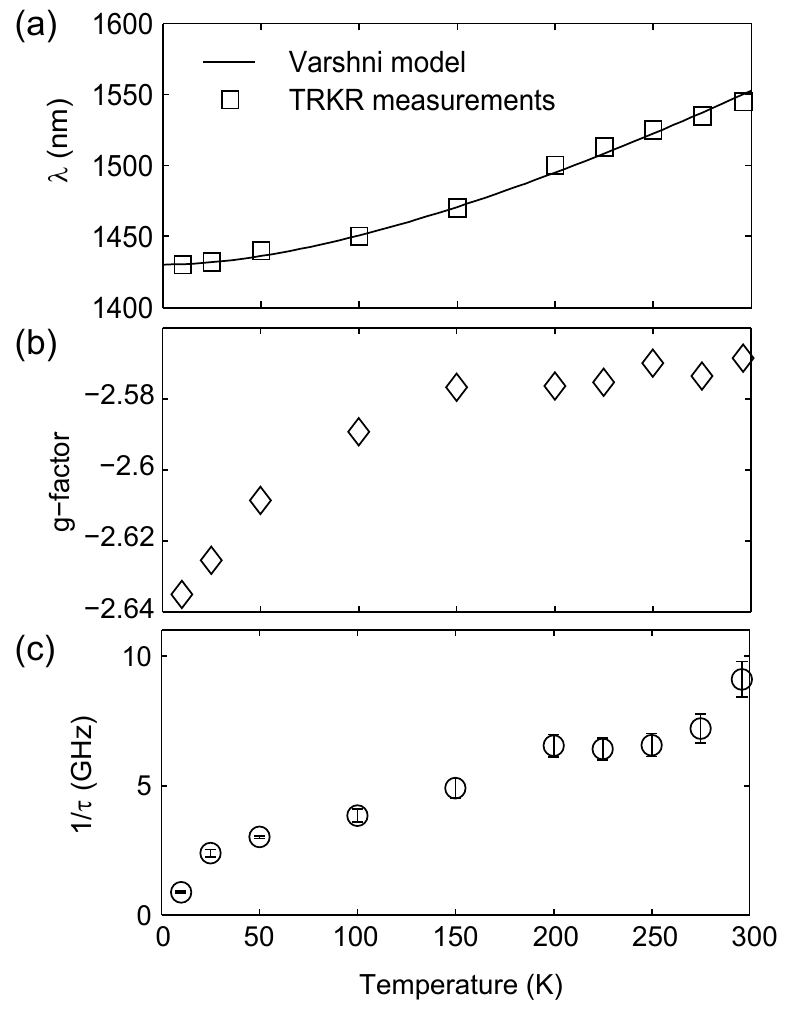}
\caption{\label{Fig2} (Color online). (a) The QW absorption edge depends on temperature; the wavelength $\lambda$ of the laser used for TRKR (symbols) is adapted accordingly. Solid line shows the wavelength predicted by a Varshni model taking parameters from InGaAs grown lattice-matched on InP. (b) The g-factor slightly increases towards positive values as $T$ is increased, similar to the dependence found in GaAs. (c) The decay rate $1/\tau$ monotonically increases with temperature. At 10\,K, the spin lifetime exceeds 1\,ns.}
\end{figure}

Figure~\ref{Fig2} summarizes the dependence of the spin dynamics on temperature $T$. As the band gap increases with decreasing $T$, we need to lower the laser wavelength $\lambda$ as we go to smaller $T$. Figure~\ref{Fig2}(a) shows the $\lambda$ used for TRKR measurements (symbols). These match the temperature dependence of a Varshni model with parameters obtained for lattice-matched InGaAs on InP~\cite{Gaskill1990}, see line in Fig.~\ref{Fig2}(a). For each $T$, we measure the electron g-factor [Fig.~\ref{Fig2}(b)]: we observe a small increase of the negative g-factor towards zero, which is in agreement with observations in GaAs that were explained by a temperature dependence of the dipole matrix elements between the conduction and valence bands~\cite{Huebner2009}. The spin decay time $\tau$ monotonically increases if $T$ is lowered, reaching more than 1\,ns at 10\,K. Figure~\ref{Fig2}(c) shows the corresponding decay rate $1/\tau$. For the sample studied, the spin lifetime is expected to be limited by the Dyakonov Perel mechanism~\cite{Dyakonov1986}, whose temperature dependence~\cite{Kainz2004} is given by an increase of the spin diffusion constant $D_{\textrm s}$ proportional to $k_{\textrm B}T$ for a non-degenerate electron gas. In addition, $D_{\textrm s}$ varies linearly with the electron scattering time relevant for spin diffusion.

\begin{figure}[t]
\advance\leftskip0cm
\includegraphics[width=0.48\textwidth]{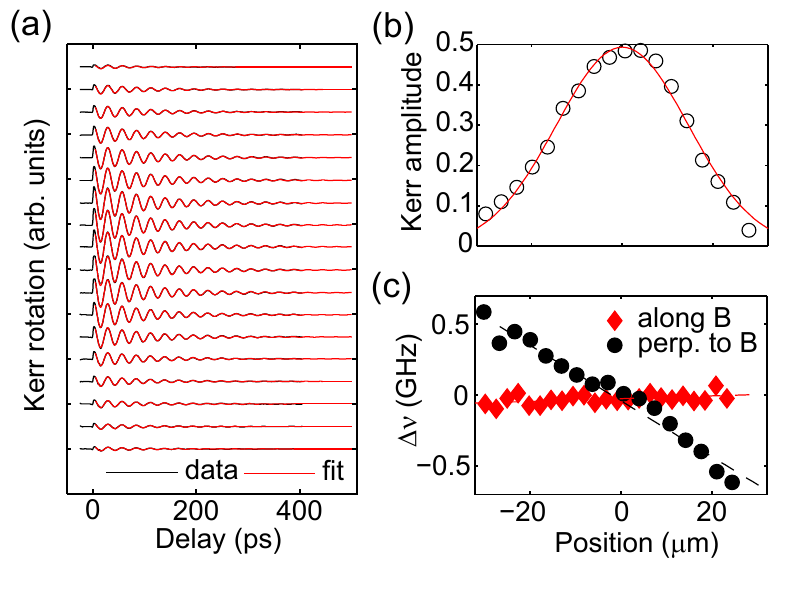}
\caption{\label{Fig3} (Color online). (a) TRKR traces taken at room temperature and $B=1.0$\,T, with the laser probe spot displaced perpendicularly to $B$ from -30\,$\mu$m (lowest curve) to 28\,$\mu$m (uppermost curve). (b) The fitted amplitude of the Kerr signal represents the convoluted pump and probe spot profiles and yields a
sigma-width of $\sigma$=14\,$\mu$m. (c) The fitted precession frequency depends on the overlap position. Shown is $\Delta\nu$, defined as half of the difference between frequencies taken at $B=\pm1$\,T. If the probe spot is displaced perpendicularly to the field, a linear change of $\Delta\nu$ is observed, whereas $\Delta\nu$ remains approximately constant if the probe beam is scanned along the field direction. The change in $\Delta\nu$ represents the addition of the effective spin-orbit magnetic field along the external field direction.}
\end{figure}

We use the technique first presented in Ref.~\onlinecite{Kohda2015} to investigate the SOI in the QWs. Electrons with a diffusive velocity $v=D_{\textrm s} x/\sigma^2$ are selected by displacing the probe pulse by a distance $x$ with respect to the position of the pump pulse. This velocity induces an effective magnetic field through the SOI, which modifies the electron spin precession frequency. The vector components of the effective spin-orbit magnetic field are determined by applying an external magnetic field $B$. Thereby, only components along the external magnetic field change $\nu$ linearly, whereas perpendicular components lead to quadratic modifications that can be neglected if the effective spin-orbit field is small compared to $B$. Figure~\ref{Fig3}(a) shows a series of TRKR traces for different pump-probe overlap positions $x$ between -30 and 28\,$\mu$m, taken at room temperature and with $B=1$\,T applied perpendicular to the overlap scan direction. As expected, the signal amplitude peaks at maximum overlap, as seen in Fig.~\ref{Fig3}(b). The Kerr amplitude $a$ is well fitted by a Gaussian profile of sigma-width $\sigma$=14\,$\mu$m. This width corresponds to the convoluted spot sizes of pump and probe beams. In Fig.~\ref{Fig3}(c), we present $\Delta\nu=(\nu(B)-\nu(-B))/2$, the difference between measurements taken with opposite directions of $B$. For overlap scans perpendicular to $B$ (black dots), $\Delta\nu$ varies linearly with the overlap position. For overlap scans along $B$, $\Delta\nu$ is flat, indicating that -- as expected -- the dominant spin-orbit field points perpendicularly to the diffusive velocity.

\begin{figure}[t]
\advance\leftskip0cm
\includegraphics[width=0.48\textwidth]{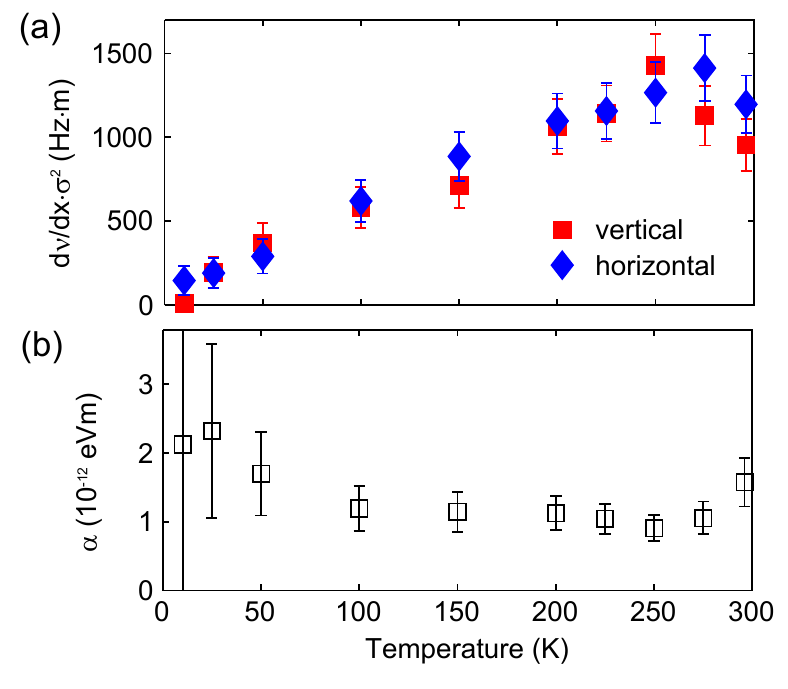}
\caption{\label{Fig4} (Color online). Normalized slopes of $\Delta\nu$ versus temperature. This quantity is proportional to the relevant spin-orbit coefficient and to the spin diffusion constant $D_{\textrm s}$. Measurements for the two sample orientations (squares and diamonds) show similar values with the same sign, indicating a predominant Rashba contribution to SOI. (b) By comparing the slope of $\Delta\nu$ with the spin decay rate $1/\tau$, values for the Rashba coefficient $\alpha$ are obtained.}
\end{figure}

In Fig.~\ref{Fig4}(a), the slopes $d\Delta\nu/dx$ are summarized for different $T$. We present the data multiplied by $\sigma^2$. A monotonic increase with $T$ is observed. The squares and diamonds correspond to the external magnetic field oriented along the sample's [110] and [1$\overline 1$0] directions. The similar values and the same sign for the two directions indicates that the Rashba SOI is much stronger than the Dresselhaus part. The Dresselhaus SOI alone would reverse the sign for the two directions~\cite{Kohda2015}, and a superposition of a significant Dresselhaus contribution to the Rashba one would change the value of the frequency slope $d\Delta\nu/dx$ for the two directions. The model presented in Ref.~\onlinecite{Kohda2015} connects the slopes with the SOI. In the Rashba-only case, we obtain $d\Delta\nu/dx\cdot\sigma^2=2D_{\textrm s}m\alpha/(\pi\hbar^2)$. Here, $m$ is the effective electron mass and $\hbar$ the reduced Planck constant. In order to get quantitative values for the Rashba coefficient $\alpha$, $D_{\textrm s}$ needs to be known. This can be obtained from the temporal expansion of the spin profile measured~\cite{Weber2005,Zhao2009,Walser2012}. In our experiment, this expansion cannot be resolved because it is too small compared with the laser spot sizes. Nevertheless, we can estimate values for $\alpha$ by taking into account the information contained in the measured values of $\tau$. Assuming a Dyakonov Perel spin relaxation mechanism~\cite{Averkiev1999,Salis2014}, $1/\tau$ is given by $6D_{\textrm s} m^2/\hbar^4\times\alpha^2$. The equation holds for both the degenerate and non-degenerate regime. The dependence on temperature enters through $D_{\textrm s}$ via the electron distribution and the dominant electron scattering mechanism. Since $1/\tau$ also depends linearly on $D_{\textrm s}$, we can remove $D_{\textrm s}$ by dividing $1/\tau$ by $d\Delta\nu/dx\times\sigma^2$. This ratio is equal to $3\pi m/\hbar^2\times \alpha$. Figure~\ref{Fig4}(b) summarizes the values for $\alpha$ obtained in this way (for the $\Delta\nu$ slopes we take the average between the two crystal directions). They represent an upper limit for $\alpha$ if one considers that additional contributions may affect the electron spin decay. We obtain values for $\alpha$ that start around 2$\times 10^{-12}$\,eVm at low temperatures and decrease to around 1$\times 10^{-12}$ eVm at room temperature. Such a large value is especially promising regarding the measured spin ensemble coherence times above 1\,ns. Considering an electric field $E$ across the QWs of about 8\,V/$\mu$m, the ratio $\alpha/E$ amounts to 25\,e{\AA}$^2$. This compares well to calculated values~\cite{Winkler2003} of $r_{41}^{6c6c}$ of 5.2\,e{\AA}$^2$ for GaAs and 117\,e{\AA}$^2$ for InAs.

In conclusion, we have demonstrated pump-probe based all-optical measurements of SOI in low-band-gap InGaAs QWs. The SOI is found to be of Rashba type with an upper limit for the coefficient $\alpha$ of 2$\times$10$^{-12}$\,eVm. These experiments can be extended to materials such as InAs, GaSb and InSb with smaller band gaps and expectedly much stronger SOI, since a spatial resolution of the spin dynamics on a scale of the spin-orbit length (expected to be much smaller than 1\,$\mu$m in these materials) is not required. These experiments will allow us to better understand and characterize the SOI in low-band-gap semiconductor quantum structures used for creating topological insulators, helical modes and Majorana fermions.

The work is financially supported by the Swiss National Science Foundation within the NCCR QSIT, and by the EU project SiSpin. We acknowledge P. Altmann and D. Widmer for experimental assistance, as well as M. Tschudy, R. Grundbacher and U. Drechsler for help with sample fabrication.

\bibliographystyle{prsty}

\end{document}